\documentclass[journal=jpcl,manuscript=letter,layout=traditional]{achemso}
\usepackage{amsmath,xcolor,graphicx,colortbl}
\usepackage{amssymb,soul}

\author{George Volonakis}\altaffiliation{These authors contributed equally to this work}
\affiliation[Department of Materials, University of Oxford]{Department of Materials, University of 
Oxford, Parks Road OX1 3PH, Oxford, UK}

\author{Marina R. Filip} \altaffiliation{These authors contributed equally to this work}
\affiliation[Department of Materials, University of Oxford]{Department of Materials, University of
Oxford, Parks Road OX1 3PH, Oxford, UK}

\author{Amir Abbas Haghighirad}\affiliation[Department of Physics, University of Oxford]{Department of Physics, 
University of Oxford, Clarendon Laboratory, Parks Road, Oxford OX1 3PU, UK}

\author{Nobuya Sakai}\affiliation[Department of Physics, University of Oxford]{Department of Physics, 
University of Oxford, Clarendon Laboratory, Parks Road, Oxford OX1 3PU, UK}

\author{Bernard Wenger}\affiliation[Department of Physics, University of Oxford]{Department of Physics, 
University of Oxford, Clarendon Laboratory, Parks Road, Oxford OX1 3PU, UK}

\author{Henry J. Snaith}\affiliation[Department of Physics, University of Oxford]{Department of Physics, 
University of Oxford, Clarendon Laboratory, Parks Road, Oxford OX1 3PU, UK}
\email{feliciano.giustino@materials.ox.ac.uk}\phone{(+44) 1865 612790 }

\author{Feliciano Giustino} \affiliation[Department of Materials, University of Oxford]{Department of 
Materials, University of Oxford, Parks Road OX1 3PH, Oxford, UK} 
\email{henry.snaith@physics.ox.ac.uk}\phone{(+44) 01865 272380 }

\title{Lead-Free Halide Double Perovskites via Heterovalent Substitution of Noble Metals.}

\begin{document}
\newpage

\begin{abstract}

Lead-based halide perovskites are emerging as the most promising class of materials for next generation 
optoelectronics. However, despite the enormous success of lead-halide perovskite solar 
cells, the issues of stability and toxicity are yet to be resolved. Here
we report on the computational design and the experimental synthesis of a new family of Pb-free
inorganic halide double-perovskites based on bismuth or antimony and noble metals. Using 
first-principles calculations we show that this hitherto unknown family of perovskites
exhibits very promising optoelectronic properties, such as tunable band gaps in the visible range and 
low carrier effective masses. Furthermore, we successfully synthesize the 
double perovskite Cs$_2$BiAgCl$_6$, we perform structural refinement using
single-crystal X-ray diffraction, and we characterize its optical properties via optical absorption
and photoluminescence measurements. This new perovskite belongs to the $Fm\bar{3}m$ space group, 
and consists of BiCl$_6$ and AgCl$_6$ octahedra alternating in a rock-salt face-centered cubic structure.
From UV-Vis and PL measurements we obtain an indirect gap of 2.2~eV. The new compound 
is very stable under ambient conditions.\\ 

\vspace{0.1cm} 

\noindent {\bf Table of Contents Image}
\begin{figure*}
\includegraphics[width=0.7\textwidth]{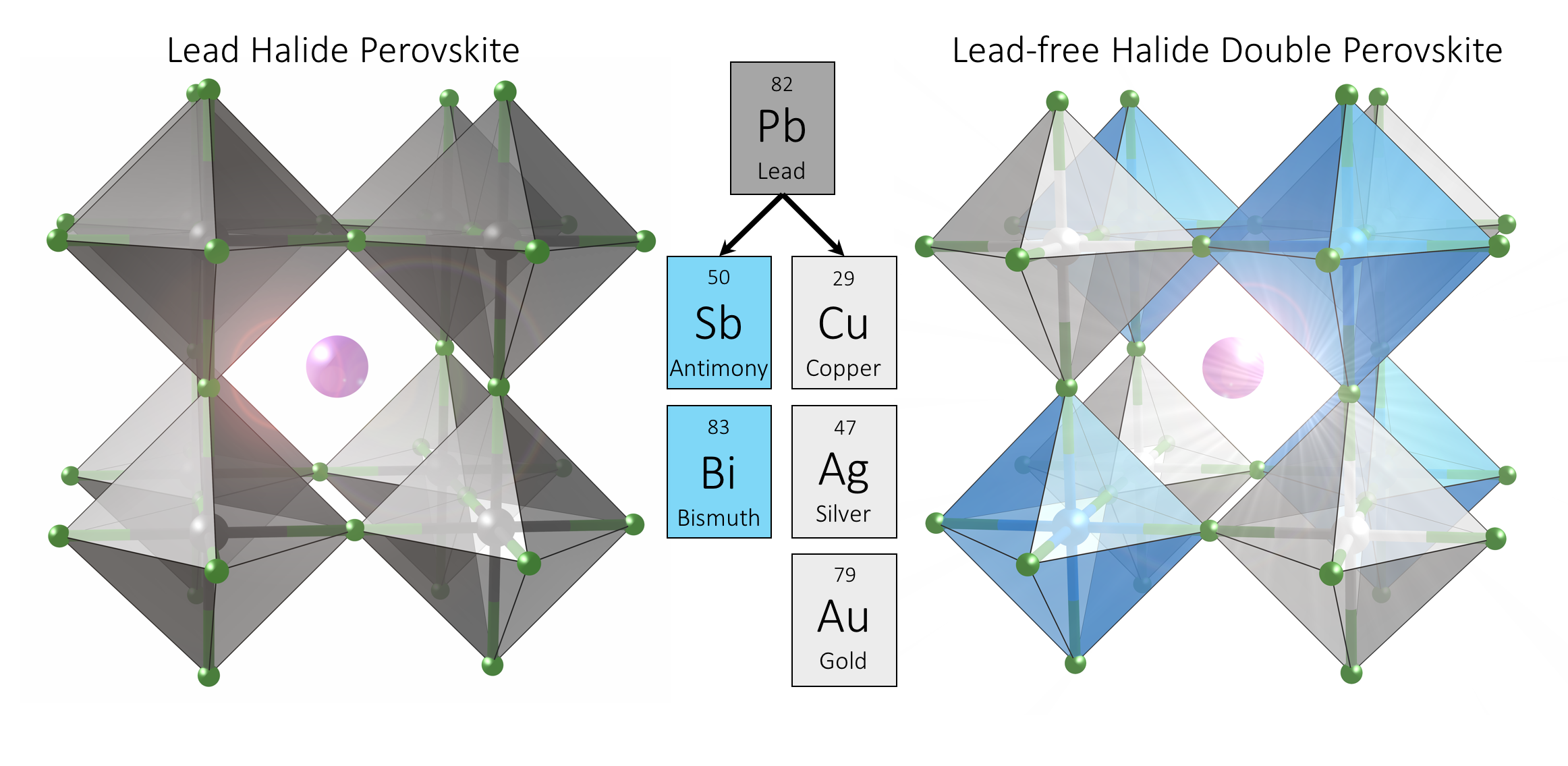}
\end{figure*}


\end{abstract}

\newpage


Perovskites are among the most fascinating crystals, and play important
roles in a variety of applications, including ferroelectricity, piezzoelectricity, high-T$_{\rm c}$ 
superconductivity, ferromagnetism, giant magnetoresistance, photocatalysis and photovoltaics~\cite{Suntivich2011, 
Stranks2015, Gratzel2014, Vasala2015, 
Grinberg2013, Ramesh2007, Rinjders2005, Ahn2004}. The majority of perovskites are oxides 
and are very stable under ambient temperature and pressure conditions~\cite{Vasala2015, Fan2015}. 
However, this stability is usually accompanied by very large band gaps, therefore most oxide perovskites 
are not suitable candidates for optoelectronic 
applications. The most noteworthy exceptions are the ferroelectric perovskite oxides related to LiNbO$_3$, 
BaTiO$_3$, Pb(Zr, Ti)O$_3$ and BiFeO$_3$, which are being actively investigated for photovoltaic applications, reaching 
power conversion efficiencies of up to 8\%\cite{Fan2015}. 
The past five years witnessed a revolution in optoelectronic research with the discovery of the organic-inorganic 
lead-halide perovskite family. These solution-processable perovskites are fast becoming the most promising materials 
for the next generation of solar cells, achieving efficiencies above 20\%~\cite{Green2014, Lee2012, 
Kim2012, NREL}. Despite this breakthrough, hybrid lead-halide perovskites are known to degrade due to moisture and 
heat~\cite{Manser2016}, upon prolonged exposure to light,~\cite{Hoke2015} and are prone to ion or halide vacancy migration, 
leading to unstable operation of photovoltaic devices~\cite{Eames2015, Meloni2016}. At the same time the presence of lead 
raises concerns about the potential environmental impact of these materials~\cite{Espinosa2015, Babagayigit2016}. Given 
these limitations, identifying a stable, non-toxic halide perovskite optoelectronic material is one of the key  challenges
to be addressed in the area of perovskite optoelectronics. 


The starting point of our search for a lead-free halide-perovskite is the prototypical inorganic compound of the family CsPbI$_3$. CsPbI$_3$ 
is an ABX$_3$ perovskite where the heavy metal cations Pb$^{2+}$ and the halide anions I$^-$ occupy the B and X sites, 
respectively, while Cs$^+$ occupies the A site. The most obvious route to replacing Pb in this compound is via substitution 
of other group-14 elements such as Sn and Ge. However both elements tend to undergo oxidation, for example from Sn$^{2+}$ 
to Sn$^{4+}$, leading to a rapid degradation of the corresponding halide perovskites~\cite{Stoumpos2013, Baikie2013, 
Hao2014, Noel2014}. More generally, it should also be possible to substitute lead by other divalent cations outside of 
group-14 elements. However, our previous high-throughput computational screening  of potential candidates showed that the 
homovalent substitution of lead in halide perovskites impacts negatively the optoelectronic properties by increasing band 
gaps and effective masses~\cite{Filip2016}. 

Another possible avenue is to consider heterovalent substitution, that is the formation of a double perovskite structure with 
a basic formula unit A$_2$BB$^\prime$X$_6$~\cite{Vasala2015}. This type of compounds are abundant in the case of oxides 
and are well known for their ferroelectric, ferromagnetic and multiferroic properties~\cite{Vasala2015}. 
Additionally, double perovskites have been explored in order to tune the band gap of oxide perovskites~\cite{Nechache2015, Berger2012}. 
On the other hand,  halide double perovskites remain a much less explored class of materials. To date, the best 
known halide double perovskites are based on alkali and rare-earth metals, and are investigated for applications as 
scintillators in radiation detectors~\cite{Loef2002}.

In order to replace the divalent Pb cations and maintain the total charge neutrality, the B$^\prime$ and B$^{\prime\prime}$ sites 
have to be occupied by one monovalent and one trivalent cation. We search for our B$^{\prime3+}$ metallic cations among the pnictogens, 
and consider Bi and Sb as the most suitable choices. Arsenic is less desirable owing to its toxicity. For the monovalent cations 
we choose the noble metals Cu, Ag and Au. From elementary considerations Cu, Ag, and Au appear very promising for optoelectronic 
applications. In fact, in their metallic form, the noble metals are the best known electrical conductors, owing to their filled 
$d^{10}$ shell and the free-electron-like behaviour of the $s^1$ shell. In addition, in an octahedral environment, the ionic radii 
of Cu$^+$ (0.91~\AA), Ag$^+$ (1.29~\AA) and Au$^+$ (1.51~\AA) are similar to those of Pb$^{2+}$ (1.19~\AA), Sb$^{3+}$ (0.76~\AA) 
and Bi$^{3+}$ (1.03~\AA)~\cite{Shannon}. Following this simple reasoning we investigate hypothetical  halide double perovskites 
with the pairs B$^\prime$/B$^{\prime\prime}$ where B$^\prime$ = Sb, Bi, and B$^{\prime\prime}$ = Cu, Ag, Au. 

\begin{figure}[t!]
\begin{center}
\includegraphics[width=0.94\textwidth]{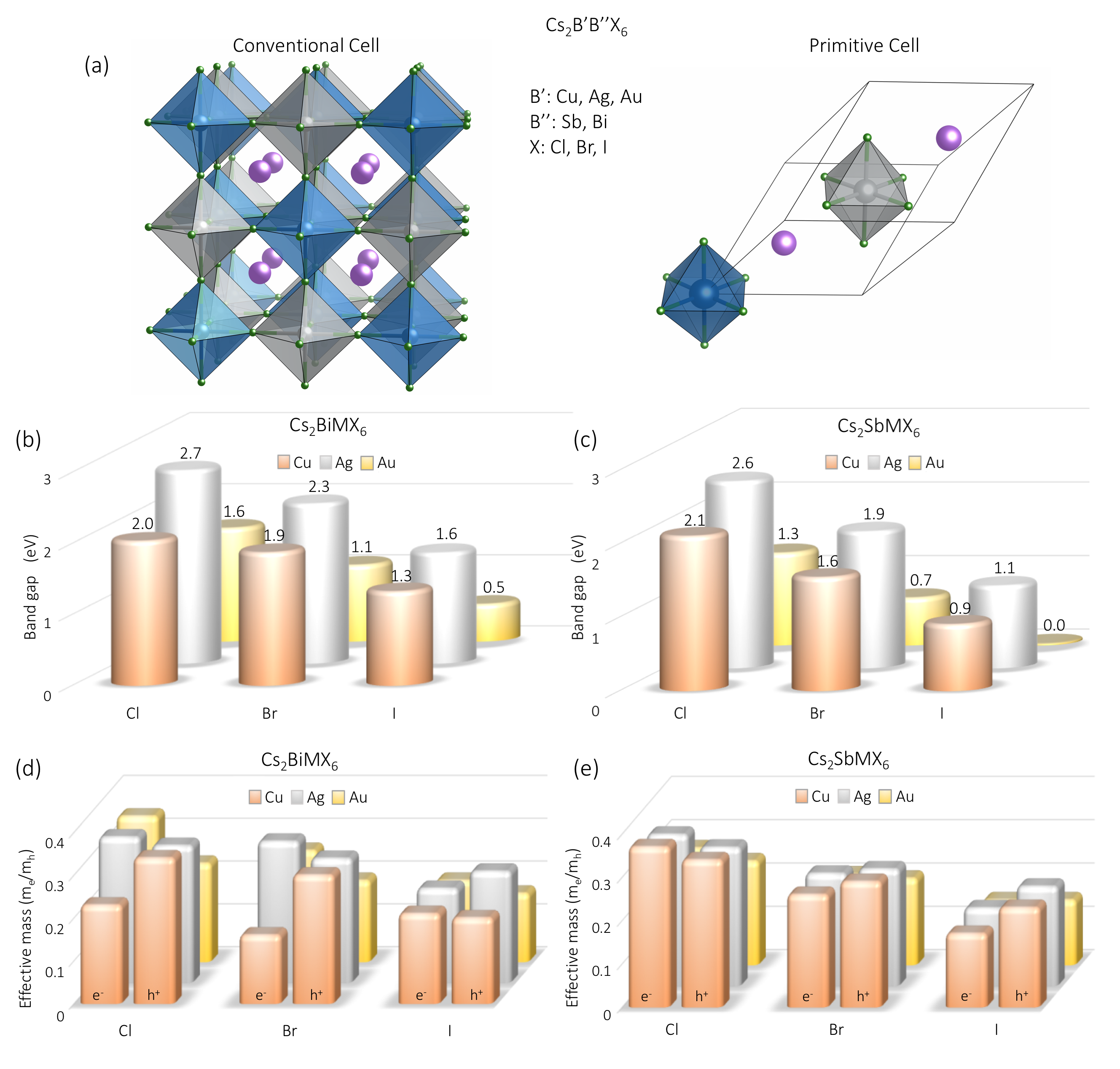}
\end{center}
\caption{\label{fig:1}
  \textbf{Computational screening of the electronic properties of the pnictogen-noble metal halide double perovskites\hspace{0.2cm}}\\ 
  \textbf{a}~Polyhedral model of the conventional (left) and reduced (right) unit cell of the hypothetical halide double perovskites. 
  The pnictogen (B$^\prime$) and noble metal (B$^{\prime\prime}$) cations alternate along the three crystallographic axes, forming 
  the rock-salt ordering.  
  \textbf{b}~Electronic  band gaps calculated for all compounds in the halide double perovskite family using the PBE0 hybrid functionals.
  All calculated band gaps are indirect with the top of the valence band at the X point (0,0,$2\pi/a$) of the Brillouin zone, where $a$ is 
  the lattice parameter of the FCC unit cell. The bottom of the conduction band is at the L point ($\pi/a$, $\pi/a$, $\pi/a$) of the 
  Brillouin zone in all cases, except Cs$_2$BiAgCl$_6$, Cs$_2$BiCuCl$_6$ and Cs$_2$BiCuBr$_6$ where the bottom of the conduction band 
  is found at the $\Gamma$ (0,0,0) point. 
  \textbf{c}~Transport effective masses calculated from DFT/LDA for each compound (see Supporting Information). The effective masses
  are calculated at the VBM (holes) and CBM (electrons)  in each case.
}
\end{figure}

We investigate the electronic properties of these hypothetical compounds from first principles using density functional theory (DFT) in the local density approximation (LDA). 
We construct `rock-salt' double perovskites, whereby B$^\prime$ and B$^{\prime\prime}$ alternate in every direction (shown 
in Figure~\ref{fig:1}a). The rock-salt ordering is known to be the ground state for most oxide double perovskites~\cite{Vasala2015}, 
therefore it can be expected to hold also in the present case. For each model structure we perform full structural optimization 
using DFT-LDA and calculate the electronic band gaps using the hybrid PBE0 functional as described in the Supporting Information.

In Figure~\ref{fig:1}b-c we show a comparative view of the band gaps calculated for the entire Cs$_2$B$^\prime$B$^{\prime\prime}$X$_6$ family. 
We find that all band gaps are below 2.7~eV, spanning the visible and near infrared optical spectrum. The band gaps are 
indirect and increase as we move up the halogen or the pnictogen column in the periodic table, but do not follow a monotonic 
trend with respect to the size of the noble metal cation. This behaviour can be explained by the character of the electronic 
states at the band edges. Indeed, a shown in Figure~S1 of the Supporting Information, the conduction band bottom and 
valence band top in each case are predominantly of pnictogen-$p$ and halogen-$p$ character, respectively. As we move up in 
the periodic table the energy of the halogen-$p$ states decreases, thus lowering the energy of the valence band top. Similarly, 
the energy of the pnictogen-$p$ states decreases when moving up in the periodic table, thus lowering
the energy of the conduction band bottom. The electron and hole effective masses calculated at the band edges exhibit an 
anisotropic behaviour in most cases (see Table~S1). Throughout the entire family of compounds the electron masses are more isotropic than the hole masses. For clarity, in 
Figure~\ref{fig:1} we report the transport effective masses~\cite{Galli2014}, as defined in the Supporting Information. 
We note that all compounds exhibit small carrier effective masses between 0.1 and 0.4~$m_e$, very close to those calculated 
for CH$_3$NH$_3$PbI$_3$ within the same level of theory~\cite{Filip2015-2}.

The electronic band structures of these  halide double perovskites (shown in Figure S2 and S3) exhibit 
several features of particular interest. In all cases, the valence band maximum (VBM) is at the $X$ (0,0,$2\pi/a$) point in the Brillouin 
zone. The conduction band minimum (CBM) is at  $\Gamma$ (0,0,0) for Cs$_2$BiAgCl$_6$, Cs$_2$BiCuCl$_6$ and Cs$_2$BiCuBr$_6$, while
for the other compounds the CBM is at the $L$ ($\pi/a$, $\pi/a$, $\pi/a$) point. The 
FCC cubic crystals of the former three compounds are indirect band gap semiconductors, however
a small disturbance of the conventional unit cell symmetry could render the direct optical transition allowed.
This is shown in Figure~S4, where the band structure of Cs$_2$BiAgCl$_6$ is calculated in the conventional unit cell 
(which corresponds to two primitive cells). Here, as a result of Brillouin zone folding in the conventional 
FCC unit cell, the band gap of Cs$_2$BiAgCl$_6$ becomes direct at the $\Gamma$ point. In practice, 
this effect could be realized by incorporating an organic cation,  like methylammonium (CH$_3$NH$_3^+$) or formamidinium 
(CHN$_2$H$_4^+$) into the cuboctahedral cavity. We illustrate this possibility by calculating the band structure
 of the hypothetical 
orthorhombic CH$_3$NH$_3$BiAgCl$_6$ (constructed as described in the Supporting Information); as expected we obtain a direct band gap, 
as shown in Figure~S5.

\begin{figure}[t!]
\begin{center}
\includegraphics[width=1.0\textwidth]{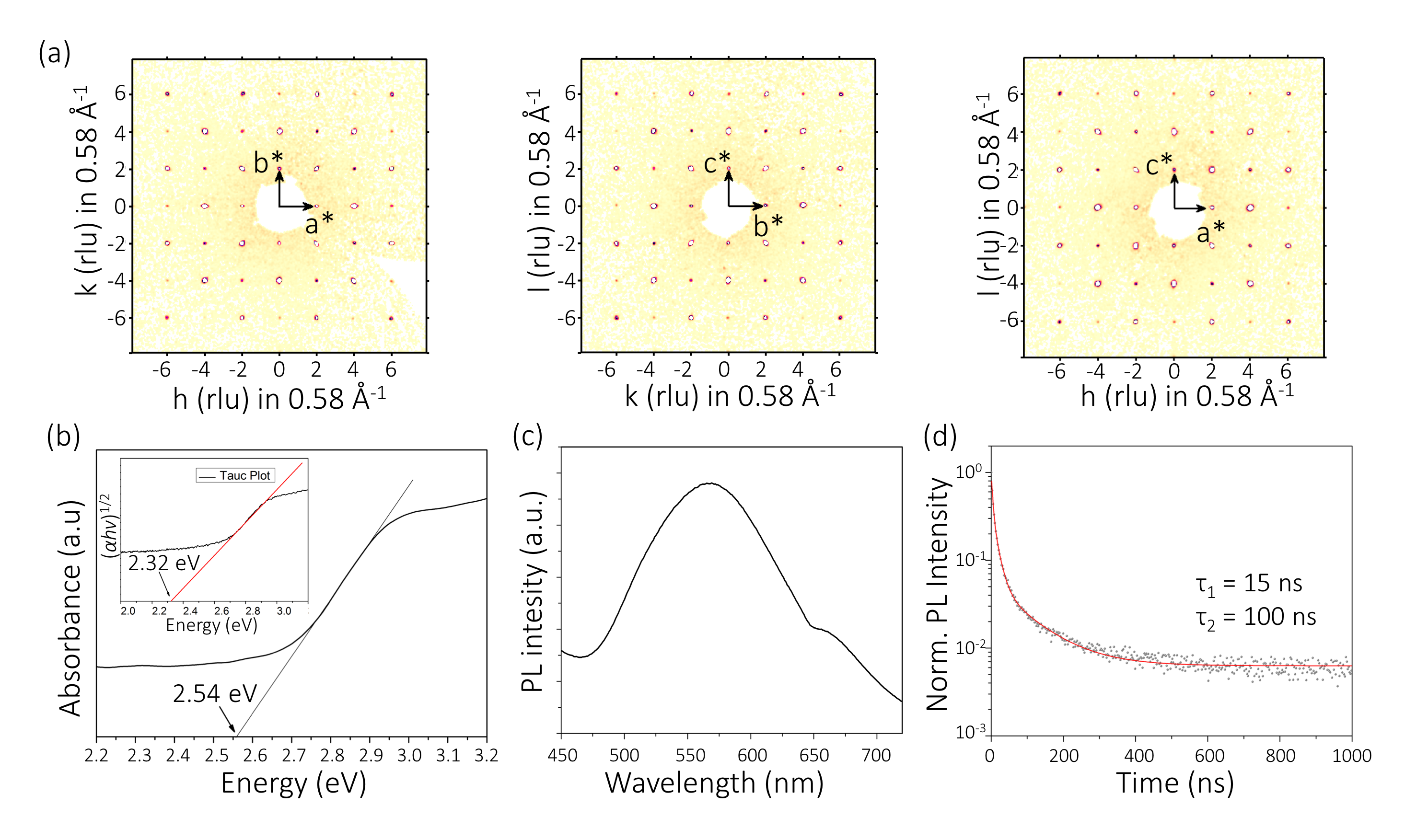}
\end{center}
\caption{\label{fig:2}
 \textbf{Experimental synthesis and characterization of Cs$_2$BiAgCl$_6$.}\\
 \textbf{a} X-ray diffraction pattern for a Cs$_2$BiAgCl$_6$ single crystal at 293 K. $hkl$ shown for three different planes,
 i.e.  $0kl$, $h0l$ and $hk0$. All wave vectors are labeled in reciprocal lattice units (rlu) and  a$^\mathrm{*}$, b$^\mathrm{*}$
 and c$^\mathrm{*}$ denote reciprocal lattice vectors of the cubic cell of the $Fm\bar{3}m$ structure.
 \textbf{b} UV-Vis optical absorption spectrum of Cs$_2$BiAgCl$_6$. The inset shows the Tauc plot, corresponding to an indirect
 allowed transition [assuming the expression: $(\alpha h\nu)^{1/2} = C(h\nu - E_g)$, where $\alpha$ is the absorption coefficient, 
 $h\nu$ is the energy of the incoming photon $E_g$ is the optical band gap and $C$ is a constant]. The straight lines are fitted to the 
 linear regions of the absorption spectrum and Tauc plot, and the intercepts at 2.32~eV and 2.54~eV marked on the plot are calculated from the fit. 
 \textbf{c} Steady-state photoluminescence (PL) spectrum of Cs$_2$BiAgCl$_6$, deposited on glass. 
 \textbf{d} Time resolved photoluminescence decay of Cs$_2$BiAgCl$_6$, deposited on glass. The data is fitted using a biexponential decay
 function. The decay lifetimes of 15~ns (fast) and 100~ns (slow) is estimated from the fit. 
}
\end{figure}

Having established that the family of  A$_2$BB$^\prime$X$_3$  halide double perovskites, based on B = Sb, Bi and B$^\prime$ = Cu, 
Ag, Au exhibits promising optoelectronic properties, we move to the synthesis and optical characterization of a representative member of 
this group of compounds. We adapt the synthesis process of Cs$_2$BiNaCl$_6$, reported in Ref.~\cite{Morss1972}, 
to allow for the incorporation of a noble metal. Of the three noble metals under consideration, Ag has an ionic radius 
which is closest to that of Na (1.02~\AA\ vs 1.15~\AA). For this reason we proceed to synthesize Cs$_2$BiAgCl$_6$ by conventional solid-state 
reaction as described in detail in the Supporting Information. In Figure~\ref{fig:2}a we show the X-ray Diffraction Pattern for 
a single crystal ($\sim$30$\mu$m  diameter). We observe sharp reflections for the crystallographic $0kl$, $h0l$ and $hk0$ planes. These reflections show 
characteristics of m$\bar{3}$m symmetry that reveal systematic absences for ($hkl$; $h+k$, $k+l$, $h+l = 2n$) corresponding to the face-centered space groups $F432$, 
$F\bar{4}3m$ and $Fm\bar{3}m$. The latter was selected for structure refinement after confirmation that Cs$_2$BiAgCl$_6$ crystallizes in an FCC
lattice. We find that there is no significant distortion of octahedral symmetry about the Bi$^{3+}$. The atomic 
positions from the structural refinement are listed in Table~S2 of the Supplementary Information. The X-ray diffraction patterns uniquely 
identify the $Fm\bar{3}m$ (no. 225) space group at room temperature, and the quantitative structural analysis gives a very good description of the data. In addition, our crystal structure refinement is  consistent with the rock-salt configuration 
assumed by our atomistic model. The experimental and computationally predicted conventional lattice parameters are in very good agreement, 10.78~\AA\ 
and 10.50~\AA, respectively.
From the optical absorption spectrum and Tauc plot (see Figure~\ref{fig:2}b) we can estimate an indirect optical band gap in the range
of 2.3-2.5~eV. The indirect character of the band gap is consistent with the broad photoluminescence peak observed between 480 and 650 nm (1.9-2.6~eV)
with the maximum at $\sim$575~nm (2.2~eV), red-shifted with respect to the optical absorption onset. In addition, the  time-resolved
photoluminescence decay shown in Figure~\ref{fig:2}c was fitted with a double exponential giving a fast component 
lifetime of 15~ns and a slow component lifetime of 100~ns.

\begin{figure}[t!]
\begin{center}
\includegraphics[width=1.0\textwidth]{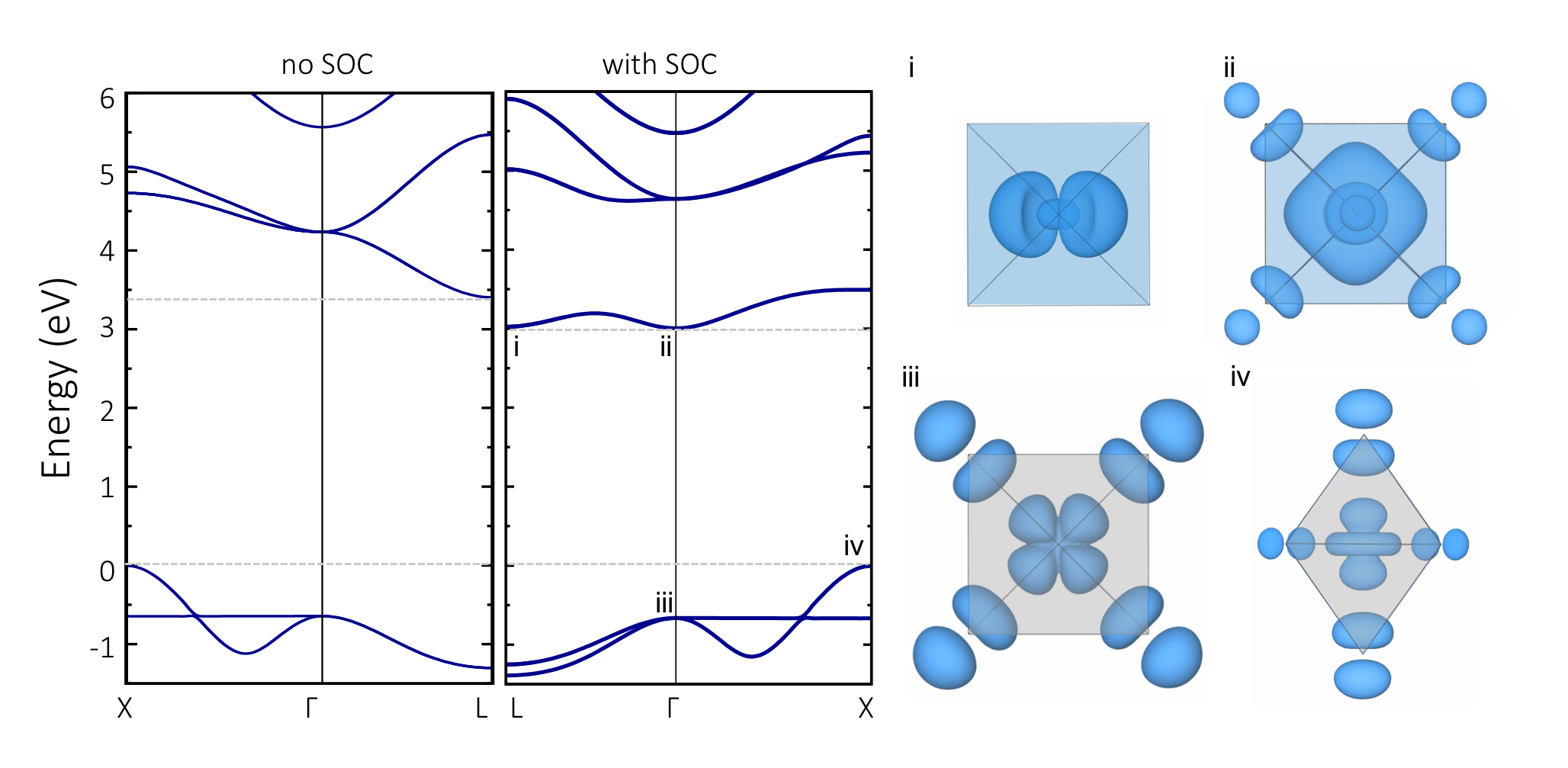}
\end{center}
\caption{\label{fig:3}
  \textbf{Electronic structure properties calculated for the experimental crystal structure of Cs$_2$BiAgCl$_6$.}\\ 
  The Band structure of Cs$_2$BiAgCl$_6$ calculated along the high symmetry path L($\pi/a, \pi/a, \pi/a$) - $\Gamma$ (0,0,0) 
  - X (0,0,2$\pi$/a) without (left) and with (right) spin-orbit coupling. 
  The black points on the fully relativistic band structure marked `i-iv' mark the conduction band bottom at L 
  and $\Gamma$ and the valence band top at $\Gamma$ and X, respectively. For each of the states we show the electronic wavefunctions. The   conduction band bottom is primarily of 
  Bi-$p$ and Cl-$p$ character, while the valence band top consists of Ag-$d$ and Cl-$p$ character. The shape of all 
  four wavefunctions is consistent with metal-halide $\sigma$-bonds. 
}
\end{figure}

In Figure~\ref{fig:3} we show the electronic band structure of Cs$_2$BiAgCl$_6$ calculated for the as determined experimental crystal structure,
with and without relativistic spin-orbit coupling effects. The features of the valence band edge are almost unchanged
when the relativistic effects are included. This is consistent with the predominant Cl-$p$ and Ag-$d$ character of this band. 
By contrast, due to the large spin-orbit coupling, the conduction band edge splits in two bands, separated by more than 1.5~eV at the $\Gamma$ point. 
This effect is not surprising, given that the character of the conduction band bottom is 
of primarily Bi-$p$ character. For comparison, in the case of Cs$_2$SbAgCl$_6$ (see Figure~S7) the 
spin-orbit splitting of the conduction band at the $\Gamma$ point is of only 0.5~eV. The fundamental band gap is reduced by 0.4~eV 
upon inclusion of relativistic effects, and the shape of the conduction band is drastically different. Therefore, the inclusion
of spin-orbit coupling is crucial for the correct description of the conduction band edge, bearing resemblance to the 
case of CH$_3$NH$_3$PbI$_3$~\cite{Even2013,  Filip2014-2}. In the fully relativistic case we calculated an indirect band gap of 
3.0~eV and lowest direct transition of 3.5~eV, in very close agreement  with the results obtained for the model Cs$_2$BiAgCl$_6$ structure, 
discussed in Figure~\ref{fig:1} [2.7~eV (indirect) and 3.3~eV (direct)]. The small difference in band gap of 0.2-0.3~eV is due to the 
small difference in volume between the experimental and predicted crystal structure. The calculated electronic  
band gaps are overestimated with respect to the measured optical band gap by approximately 0.5~eV. This quantitative discrepancy does not 
affect the qualitative physical trends of the band gaps discussed throughout this work, and can be associated to the approximations 
employed in our PBE0 calculations. A better agreement with experiment can be reached by fine-tuning the fraction of exact 
exchange, or by performing $GW$ calculations~\cite{Hedin, Hybertsen}. The latter will be reported in a future work. 

In summary, through a combined theoretical and experimental study, we have designed a new family of halide double-perovskite 
semiconductors based on pnictogens and noble metals. These compounds have promising electronic properties, such as low carrier effective 
masses and band gaps covering the visible and near-infrared region of the optical spectrum. All compounds are indirect gap 
semiconductors and exhibit strong spin-orbit coupling. 
We successfully synthesized Cs$_2$BiAgCl$_6$, and obtained a face-centered cubic double perovskite, exhibiting optical 
properties consistent with an indirect gap semiconductor, in agreement with our computational predictions. 
The present work is the first detailed description of the structure and optoelectronic properties of the pnictogen-noble metal 
halide double perovskite family, and calls for many future experimental and theoretical studies in order to assess the full 
potential of these new materials. We expect that a complete mapping of the genome of halide double perovskites based on 
pnictogens and noble metals may unlock a world of new exciting optoelectronic materials for solar cells, photodetectors, 
light-emitting devices, and transistors.

\begin{acknowledgement}
The research leading to these results has received funding from the the Graphene Flagship 
(EU FP7 grant no. 604391), the Leverhulme Trust (Grant RL-2012-001), the UK Engineering and Physical 
Sciences Research Council (Grant No. EP/J009857/1 and EP/M020517/1), and the European Union Seventh Framework 
Programme (FP7/2007-2013) under grant agreements n$^\circ$239578 (ALIGN) and n$^\circ$604032 (MESO).
The authors acknowledge the use of the University of Oxford Advanced 
Research Computing (ARC) facility (http://dx.doi.org/
10.5281/zenodo.22558) and the ARCHER UK National 
Supercomputing Service under the `AMSEC' Leadership project. G.V., M.R.F. and F.G. would like to thank Marios 
Zacharias for useful discussions. Figures involving atomic structures were rendered using VESTA \cite{VESTA}.
\end{acknowledgement}

\vspace{0.6cm} 

\noindent
 {\large \bf Note added}

\noindent 
During the preparation of this manuscript we became aware of the publication of two related papers: Ref.~\cite{Slavney2016} (published
February 7th, 2016) and Ref.~\cite{McClure2016} (published February 10th, 2016). The key difference between the present
work and that of Ref.~\cite{Slavney2016, McClure2016} is that we perform a computational screening of the entire family 
of pnictogen-noble metal double halide perovskites and perform experiments that confirm our predictions. 
\bibliography{bibliography}

\end{document}